# An Overview of Machine Learning Approaches in Wireless Mesh Networks

Samurdhi Karunaratne and Haris Gacanin
Nokia Bell Labs, Antwerp, Belgium

*Abstract*—Wireless Mesh Networks (WMNs) have been extensively studied for nearly two decades as one of the most promising candidates expected to power the high bandwidth, high coverage wireless networks of the future. However, consumer demand for such networks has only recently caught up, rendering efforts at optimizing WMNs to support high capacities and offer high QoS, while being secure and fault tolerant, more important than ever. To this end, a recent trend has been the application of Machine Learning (ML) to solve various design and management tasks related to WMNs. In this work, we discuss key ML techniques and analyze how past efforts have applied them in WMNs, while noting some existing issues and suggesting potential solutions. We also provide directions on how ML could advance future research and examine recent developments in the field.

## I. Introduction

IEEE 802.11 (Wi-Fi) network access has become so ubiquitous in recent years that one expects such connectivity everywhere, whether at home, workplace, restaurant or plane. Due to their poor coverage and low QoS guarantees, single access point (AP) networks have failed to meet increasing broadband service requirements, resulting in a demand for multi-AP networks called Wireless Mesh Networks (WMNs) [1].

A WMN generally consists of mesh gateways (MGs), mesh routers (MRs), mesh clients (MCs) and a set of wireless links among them, as illustrated in Fig. 1(a). An MC can be regarded as a user device and is in most cases, an end-point of a flow of traffic through the network. The MCs are serviced by a wireless backbone formed by the MRs. The MGs act as the points at which a WMN is connected to wired infrastructure, and typically, to the Internet. Therefore, a network request originating at an MC would be transferred through its associated MR onto the wireless backbone, where it takes one or more hops to reach an MG before reaching the Internet (and vice versa).

Several factors affect the service (e.g. throughput, delay) experienced by an MC in a WMN, such as interference from other signals and contention due to simultaneous transmissions, to name just a few. To obtain a demanded level of service, various design challenges like channel allocation, routing, resource allocation and deployment strategy should be addressed, paying special attention to the intricacies that each problem entails.

Rule-based deterministic techniques that were initially introduced to solve these challenges produce satisfactory performance guarantees, but lack robustness in the face of an ever-changing network environment. Thus, real-time optimization algorithms need to be adaptable to adjust themselves to recover from lost performance. Machine learning (ML) techniques are a fitting match to this description, as they can deduce the best decisions to be made by analyzing their growing database of past network statistics and performance data.

The core objective of ML is to improve the performance of a system carrying out a set of tasks by statistically analyzing the data it has gathered during the execution of previous tasks. ML techniques have been typically classified as supervised, unsupervised and reinforcement learning [2]. Supervised learning happens when the input data to a learner is already labeled with human driven guidance of the learning agent. The input data to an unsupervised learning agent is unlabeled, so the learner must identify features or patterns in the dataset to label the data by itself. As such, unsupervised learning reflects true artificial intelligence more closely, and is generally more complex than supervised learning. Reinforcement learning (RL) is another type of ML technique where the learner perceives its environment to incrementally conduct actions that try to maximize the cumulative value of a reward given in response to previous actions.

There has been increasing interest in the application of ML in WMNs over the last decade. These attempts have been directed at optimizing various aspects of WMNs to improve user throughput, reduce end-to-end delay or satisfy other QoS demands, while also trying to improve reliability, fairness and security. This paper provides readers with a comprehensive overview of the application of different ML techniques in solving major functional design problems and handling management-level tasks in WMNs. We also classify and expound on these techniques to accentuate the problems in WMNs which could characteristically be solved by them. Our conclusions and future directions are given at the end.

## II. Applications of ML for Functional Design Problems in WMNs

When designing a WMN, various functional challenges that determine the performance of the network need to be effectively addressed; ML has aided in this by becoming an invaluable decision-making tool. In each following sub-section, we explore a specific design problem and conclude our discussion with a summary highlighted in Table I.

### A. Routing

Routing is essentially deciding which route—among many possible ones—to take towards the destination at each intermediate MR along the path from source to destination. The literature is rich in approaches tackling the routing problem in

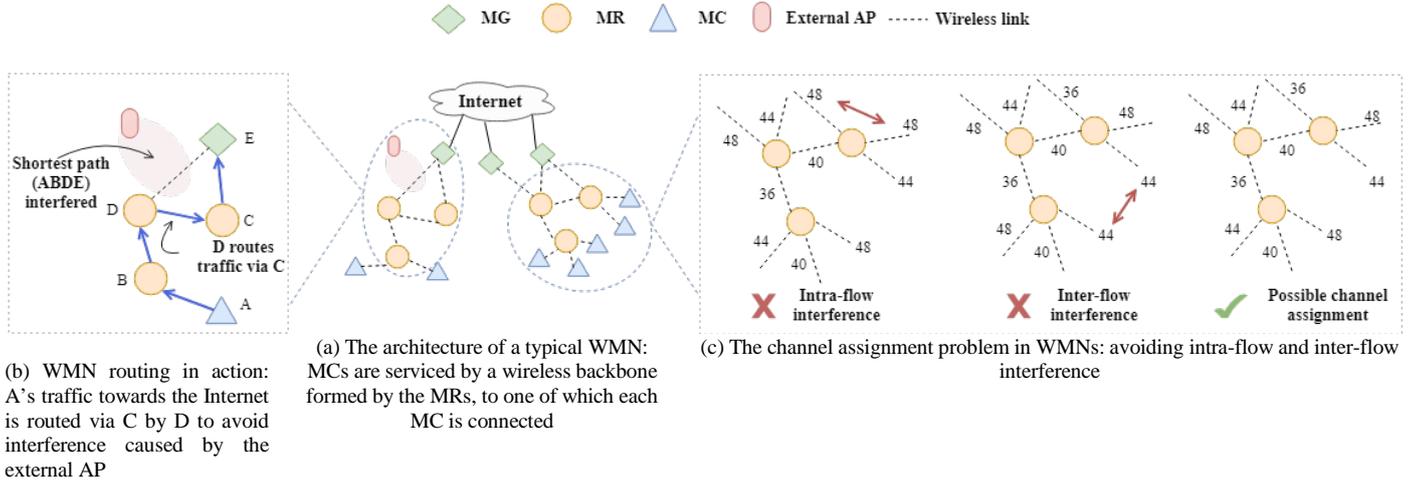

(b) WMN routing in action: A's traffic towards the Internet is routed via C by D to avoid interference caused by the external AP

(a) The architecture of a typical WMN: MCs are serviced by a wireless backbone formed by the MRs, to one of which each MC is connected

(c) The channel assignment problem in WMNs: avoiding intra-flow and inter-flow interference

**Figure 1**. Overview of a WMN and its design issues: MG = Mesh Gateway, MR= Mesh Router, MC=Mesh Client

WMNs using ML, with reinforcement learning (RL) based models being extensively used. In most practical applications of RL, the learning agent is biased, as depicted by Fig. 2(a). The bias represents some domain-specific knowledge used to guide the learning agent towards convergence.

Fig. 1(b) depicts an instance of a typical routing algorithm in action: due to link DE being interfered by the external AP, MR-C routes MC-A's traffic towards the Internet via MR-C on a longer (4-hop) path, which may otherwise have been sub-optimal. RL lends itself nicely into the routing problem, as in each routing decision, these routes could be tried out in a fashion similar to *trial-and-error* (the bias, in this case, could simply be the elimination of a few theoretically ineffective routes). It has become common practice to perform the learning in a distributed fashion, where each MR learns the best routing decisions to be made for itself, without considering other MRs. The learning can be made on a per-hop or per-flow basis; that is, the traffic source can decide on the entire path to be taken or each MR can decide only on the next hop to be taken, towards a given destination. The destination here is the Internet in most cases, so when there are multiple MGs available, the traffic source must intelligently decide on a target MG as well.

**Q-learning**: Q-learning has been amply used in WMN routing, to good effect. In Q-learning, there is a Q-value $Q(s, a)$ associated with performing action $a$ at a state $s$ that is updated each time that action is performed. At a given state, the action with the largest cumulative Q-value is considered the optimal action. Here, a key factor in deciding the convergence time is how the compromise between exploration (selecting a non-optimal decision) and exploitation (selecting the best decision so far) is struck. This compromise may be made in different ways including simple greedy, epsilon greedy and soft-max [2].

The most common routing strategy is to guide (i.e. bias) the RL-agent by facilitating it to estimate the best path based on a rule-based mechanism using certain metrics or physical parameters. For example, in [3], the authors introduced a distributed algorithm called RLBPR where an RL-agent in each MR learns the best neighbor to send an incoming packet towards a given MG. While using an epsilon greedy Q-learning strategy, each MR also makes use of theoretical estimates of the best path to the given gateway by calculating a parameter called Path Quality (PQ), for each possible path. This represents bias in the RL scheme.

**Learning Automata (LA)**: LA based methods have also been used extensively for WMN optimization tasks. The environment of a learning automaton can be described by a set of possible actions $\mathcal{A}$, a set of inputs $\mathcal{I}$ and a set of penalties (or rewards) $\mathcal{R}$ corresponding to each action. The automaton maintains a probability vector ($\psi$) which represents the probability that any action could be selected. Once an action is selected, if a penalty is received, the probabilities for all the other actions are increased and that for the selected action is decreased. One striking difference from Q-learning is that not only the probability of the selected action is affected in LA; every action is affected. LA are suited for distributed decision making in highly stochastic environments.

A multicast routing protocol called Learning Automata based Multicast Routing (LAMR) uses LA installed on each interface of a node to build a multicast tree from minimal end-to-end delay paths between the source and each multicast receiver [4]. Then, the LA optimize the initial tree to get a minimal interference tree. In [5], another multicast routing algorithm called Distributed Learning Automata-based Multicast Routing Algorithm (DLAMRA) is proposed. It shrinks the action set of an MR by constructing a minimum Steiner Connected Dominating Set (SCDS) iteratively, using LA distributed in each node. Also, the learned information is distributed among neighboring nodes to increase the convergence rate.

We should also make note of one specific issue affecting the use of RL techniques like Q-learning and LA in WMN routing—the delay of collecting feedbacks. In contrast to a design problem like channel allocation, routing decisions need to be made at a much greater frequency. Therefore, collecting feedbacks for each and every decision may not be feasible, primarily due to two reasons:
1) increased control overhead that might lead to link congestion
2) delayed update of the database causing the approach to be less reactive (especially in highly dynamic environments).

One possible solution is to only give a single collective reward for a batch of consecutive actions, instead of one for each action. Although being less granular, this has the added benefit of mitigating oversensitivity of the RL-agent due to transient changes in the environment. This mechanism is portrayed in Fig. 3.

**Artificial Neural Networks (ANNs)**: ANNs have been developed to mimic the operation of a human brain, to mostly aid in recognizing non-linear relationships in datasets. An ANN usually consists of nodes called artificial neurons. They compute output(s) based on a non-linear function of its inputs, which may be originating from other nodes or be external inputs. A single-layer ANN is portrayed in Fig. 2(b).

A multicast routing algorithm using a type of ANN called Cerebellar Model Articulation Controller (CMAC) has been proposed to predict the probability of route and node disconnection (failures) to assist in selecting better routes [4]. The input space of the CMAC is quantized into discrete states called blocks, and memory cells will be associated with each state to store information (output) for that state. CMAC neural networks exhibit advantages like speedy learning and exceptional convergence properties.

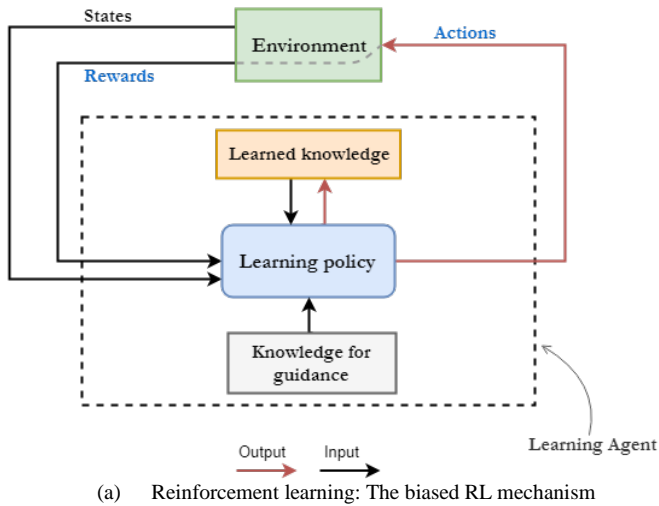

(a) Reinforcement learning: The biased RL mechanism

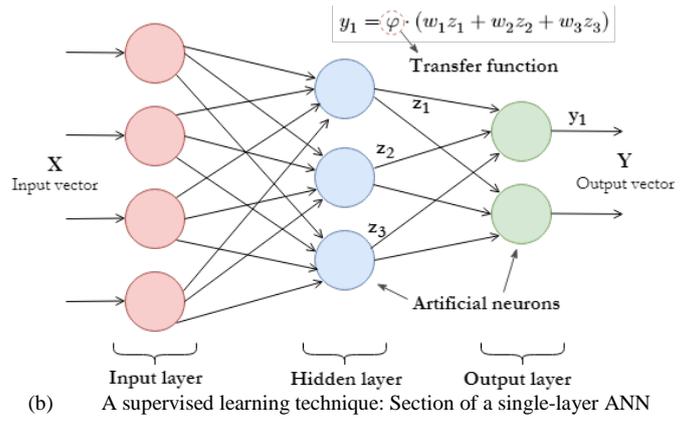

(b) A supervised learning technique: Section of a single-layer ANN

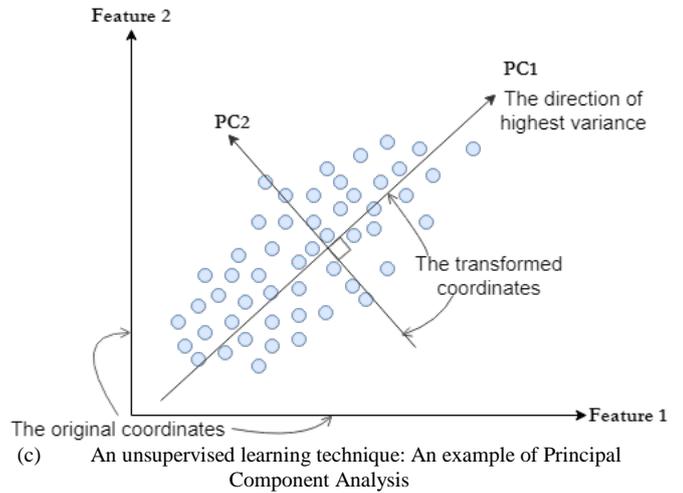

(c) An unsupervised learning technique: An example of Principal Component Analysis

**Figure 2**. Visual depictions of selected ML techniques used in Wireless Mesh Networks

## B. Channel Allocation

The channel allocation problem deals with allocating channels to the wireless links among MRs in a WMN such that interference effects are minimized and channel utilization is maximized. For example, consider the scenario given in Fig. 1(c): four non-overlapping channels 36, 40, 44 and 48 (in the 5 GHz band) should be assigned among the given links to avoid the possibility of, 1). intra-flow interference (between links of the same flow) and 2). inter-flow interference (between links of adjacent flows). It is clear that only the last assignment satisfies both requirements (other links with the same channel are too far apart to interfere).

A fair share of ML related WMN channel allocation approaches in the literature employ LA based strategies. Most of these are distributed in nature, with each of the MRs acting as a learning automaton. To this end, Leith and Clifford [6] proposed a self-managed LA based algorithm that does not require any communication between MRs. Each automaton maintains and updates a vector $\psi$, which contains a probability corresponding to each channel that reflects its history of interference. If the current channel quality is above a certain threshold, the MR will continue to operate in it; otherwise, a channel is selected randomly based on the current value of $\psi$.

They also theoretically proved that the convergence of their algorithm is guaranteed, provided that the channel assignment was feasible.

**Bayesian learning**: Bayesian learning tries to calculate the posterior probability distribution of the target features of a testing object conditioned on its input features and the entire training data set. An example for an object could be a wireless channel, while its features could be measurement data on its signal, noise and intereferenece levels measured at a radio operating on that channel at a particulat MR. Bayesian learning is well suited for occasions where there is a limited number of data points and when outliers need to be handled well. Examples include Maximum Likelihood Estimation (MLE) and Maximum A Posteriori Estimation (MAP).

In the practical application of Bayesian models, Gibbs sampling provides a convenient way to approximate posterior distributions. Authors of [7] proposed a self-organized method for automatic channel assignment in IEEE 802.11 WLANs with the aim of minimizing the total interference received by all MRs, using a Gibbs sampler. For this, they define an energy function on each node where the energy depends on the channel assignment to that node. This is utilized in an iterative procedure where the network converges to a collection of states with minimum global energy—the optimal channel assignment.

**k-means clustering**: k-means clustering groups a set of unlabeled data consisting of $n$ observations into a group of $k$ clusters based on the similarity of a set of features that are provided. Each cluster has a centroid, which can be used as it's label and is usually defined as the mean of the data points within that cluster. The most common algorithm for clustering uses an iterative refinement technique.

While k-means clustering has been applied on several classification and decision-making tasks related to WMNs in the literature, we want to highlight one potential application of this technique related to channel assignment. Several algorithms using rule-based procedures cluster nodes to several groups with the purpose of treating channel allocation in a divide and conquer fashion. A typical approach would assign the same channel to radios within the same cluster and would present a methodology to assign channels to radios on the boundary between clusters to achieve inter-cluster connectivity. We note that k-means based clustering could be used to intelligently cluster the set of MRs for this purpose. Most such approaches also require a cluster head to function; k-means clustering is naturally suited for this as the cluster centroids output by it could be used for this purpose (or some variation of it).

### C. Network Deployment

Network deployment typically deals with placing the MGs and MRs at locations which are optimal to achieve maximal network performance. Even though many other optimization problems like routing and channel assignment assume a pre-defined placement of these nodes, the performance of their ultimate outcome is dependent upon the initial physical arrangement of nodes.

MGs are extremely important in a network where most traffic is destined towards the Internet, like in a home network. More gateway nodes are beneficial as it would result in shorter paths on average for most MCs. However, this is unrealistic in most real-life deployments where the number of different wired connections to the Internet is extremely limited (in a home network this is rarely more than one). So, the problem is the compromise of figuring out the minimum number of gateway nodes and the ideal location for them to be placed. The placement of MRs is critical as it determines network coverage. It is equally important in channel assignment—in some occasions, the locations could be in such a way that no channel assignment is likely to improve performance beyond a satisfactory level. For example, the MRs could be located in areas of high interference caused by neighboring external APs; relocation of MRs is essential in such cases.

Metaheuristic techniques like Simulated Annealing (SA), Genetic Algorithms (GA) and Particle Swarm Optimization (PSO) have virtually become the *de facto* standard for intelligently solving MR and MG placement problems in WMNs. Since these techniques fall more under the umbrella of Artificial Intelligence rather than ML, we will not discuss them in detail. However, it must be noted that RL techniques like Q-learning and LA may still be worth exploring in this regard. A recent attempt has been made successfully at solving the MR placement problem where the idea of balancing the fronthaul and backhaul throughput of an MR was employed as a strategy [8]. Semi-supervised support vector machines (S3VMs), which are a variation of support vector machines (see SVMs, section III-C) that support unlabeled data, were used to identify throughput regions, while an exploration and exploitation strategy like RL was used as the learning strategy.

TABLE I. Summary of WMN design problems and ML techniques as typical solution tools

| Design problem | Objective | ML techniques used |
|---|---|---|
| Routing | Figuring out the lowest cost path to direct traffic from a source node to a destination node | ANN [4], Q-Learning [3], LA [5], MDP [10] |
| Channel assignment | Assigning channels to radio(s) of nodes with minimum effects of interference | Bayesian learning [7], LA [6] |
| Network deployment | Placement of MGs and MRs to meet network demands like coverage | SVM [8] |
| Rate adaptation | Rate at which data is transmitted between each pair of nodes in a flow | Q-Learning [9], LA [9] |
| Joint approaches | Solving multiple design problems in a complimentary manner | MDP [10] |

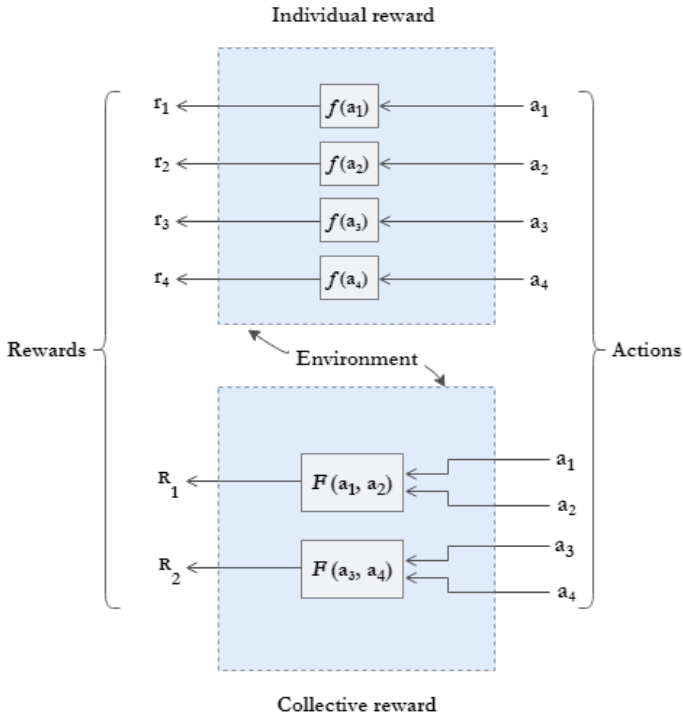

**Figure 3**. Combating delayed feedbacks in RL using collective rewards

*D. Rate adaptation*

A WMN must expect a number of different flows of traffic at any given moment (one of the primary objectives of a WMN is supporting a higher number of simultaneous users than a single AP network). From a perspective of improving per-user throughput and fairness, it is vital that these transmissions between source-node pairs happen concurrently. In conjunction with scheduling these different flows of traffic, the data rate at which they are transmitted is of importance as it ultimately impacts the inter-flow conflicts and hence, throughput.

A good variety of ML techniques has been employed for the rate adaptation problem. Stochastic Automatic Rate Adaptation Algorithm (SARA) uses a Stochastic Learning Automata (SLA) based mechanism to adjust the probability of each rate [9]. It sets equal probability for each rate in the beginning, chooses a data rate to transmit according to the current probability and updates the probability vector according to subsequent throughput values corresponding to each rate. Another method takes a Bayesian approach by using the MCS (Modulation and Coding Scheme) with a minimum estimated transmission time, where the estimation is done by using a Maximum Likelihood Estimator (MLE) [9].

*E. Joint approaches*

All the above design challenges are in fact sub-problems of the singular problem of designing a WMN where all of them are optimized to co-exist and more importantly, complement each other. In a real-life WMN, we ultimately need to use solutions to these problems together, and even though they may perform well individually, they might act in detrimental ways to one another when deployed together. For example, the physical arrangement of the network restricts the improvements that could be made with channel assignment, which together with power control determines the connectivity. Decreased connectivity limits the number of possible routes in the network.

**Markov Decision Process (MDP)**: An MDP is just like a Markov Chain, except the transition matrix depends on the action taken by the decision maker (agent) at each time step. The agent receives a reward, which depends on the action and the state. The goal is to find a function, called a policy, which specifies which action to take in each state, so as to maximize some function (e.g., the mean or expected discounted sum) of the sequence of rewards. One can formalize this in terms of Bellman's equation, which can be solved iteratively using policy iteration. The unique fixed point of this equation is the optimal policy.

In [10], Zhang et al. proposed a joint admission control and routing protocol that provides QoS guarantees in WMNs based on IEEE 802.16. The problem was modeled as a semi-Markov Decision Process (SMDP) and solved using a linear programming-based algorithm. The actions of the SMDP framework were whether or not to admit a user when a new or handoff connection request arrives, and to which route the incoming connection should be assigned. Multiple service classes were prioritized by imposing a different reward rate for each service class (service classes are defined in IEEE 802.16). The action chosen was based on the number of sessions of each class of traffic.

TABLE II. Summary of a few WMN design problems and ML techniques used to solve them.

| Management problem | Objective | ML techniques used |
|---|---|---|
| Anomaly and intrusion detection | Detecting and alerting users about possible attacks | DT & SVM [11], Perceptron [12] |
| Integrity and fault detection | Identification of faults and/or changes in the network | PCA [13] |
| Performance analysis | Studying effect of network parameters on its performance | SVM [14] |
| Fairness improvement | Striving to balance user experience among users | MDP [15] |

### III. APPLICATIONS OF ML FOR NETWORK MANAGEMENT IN WMNs

When maintaining a WMN, it is critical to pay attention to certain management-level issues that may compromise the security, integrity or the expected performance level of the system. As network demands, computing protocols and user expectations have become more and more complex over the years, ML has proved to be a vital instrument in developing tools to meet these challenges. A summary of ML applications for management tasks in WMNs is given in Table II.

*A. Anomaly/Intrusion Detection and other Security Issues*

Intrusion Detection Systems (IDS) are used to alert the users about possible attacks, ideally in time to stop the attack or mitigate the damage. They consist of three functions: 1) Event monitoring: The IDS must monitor some type of events and maintain the history of data related to these events. 2) Analysis engine: The IDS must be equipped with an analysis engine that processes the collected data to detect unusual or malicious behavior. 3) Response: the IDS must generate a response, which is typically an alert to system administrators.

**Decision Tree (DT)**: DTs are learning trees where the internal (non-leaf) nodes represent decision conditions and the leaf nodes represent a class or a feature of the input object (depending on whether a classification or a regression is being performed). By iterating down the tree, a final decision can be made. A number of different algorithms like C4.5 and ID3 can be used to construct decision trees from class-labeled training tuples.

In [11], a cross-layer based IDS is presented which trains a normal profile from features collected from both the MAC layer and network layer. It includes four components: data collection, profile training, anomaly detection and alert generation. Raw data sets are processed and loaded into the profile training module in which they used several classifiers like C4.5 (DT) and SVM (described in sub-section III. C) for pattern learning. Finally, any observed behavior that deviates significantly from the profile is considered an anomaly and an alert is triggered.

Due to the openness of the medium and the broadcast nature of transmissions, routing protocols used in a WMN should be secured against attacks that try to exploit these characteristics. To that end, either new routing protocols for WMNs that take this factor into consideration should be developed, or else existing routing protocols should be modified to address this issue specifically. Most approaches of the latter type present extensions to the Ad hoc On-Demand Distance Vector (AODV) and Hybrid Wireless Mesh Protocol (HWMP) protocols.

**Perceptron**: The perceptron is a supervised learning algorithm for binary classification (functions that can decide whether an input belongs to some specific class or not). It is a type of linear classifier that allows for online learning, in that it processes elements in the training set one at a time (the entire training set is not required in advance). It makes its predictions based on a linear predictor function combining a set of weights with the feature vector.

A distributed IDS designed to detect malicious route floods in a WMN running the AODV routing protocol is presented in [12]. Each MR intercepts the incoming traffic and gathers a set of statistics for each of its neighboring MRs. These statistics are fed as input to a perceptron classifier, which gives a binary output: a positive output is indicative of a normal condition and a negative output is indicative of an attack. To train the perceptron, $N$ training examples consisting of normal and attack instances are presented to an adaptive algorithm iteratively, which is repeated until all the examples are correctly classified.

*B. Integrity and Fault Detection*

The inherent features of wireless communication such as interference, limited bandwidth, packet loss, dynamic obstacles, and propagation loss make WMNs unstable and somewhat unreliable in certain occasions. As such, they may experience various failures, for example, node or link failures which may result in service interruption or degradation of performance. Hence, it is crucial to develop methods that can monitor the network and identify faults accurately and descriptively so that they can be remedied quickly.

**Principal Component Analysis (PCA)**: PCA takes a set of data and tries to reduce it into several principal components (PCs), which is a set of linearly uncorrelated variables. These components are ordered by the variance in the data that each component encapsulates, such that the first component has the highest variance. Being an exercise in finding maximal variance, prior normalization of input data is of utmost importance in PCA. It has applications in any problem where the number of variables is too large for a computationally feasible solution to be achieved and a smaller subset of those variables needs to be considered. The simplest case of PCA is depicted in Fig. 2(c).

PCA has been used in the fault detection of WMNs in [13]. Here, they analyze the number of packets transmitted in $l$ flows of data measured in $p$ successive time intervals. After the application of PCA, they are able to differentiate flows with high variance (abnormal flows), and hence, identify faults. By developing an identification scheme which involves reverse-mapping the derived principal components back into the measurement space, they were not only able to reduce the number of false alarms but also to pinpoint the nodes causing the actual anomalies.

*C. Miscellaneous applications*

1) *Performance analysis*

To gain a deeper understanding of the capabilities of WMNs, it is important to study the effect that different network parameters have on network output, measured in terms of a variety of QoS metrics.

**Support Vector Machines (SVMs)**: In an SVM, each data point is represented as an $n$-dimensional vector and the goal is to construct hyperplanes that best separate the set of data points into classes. The best separation can be defined in many ways: one being the hyperplane that maximizes the distance to the nearest data point of any class.

The impact of topology characteristics like number of 1-hop neighbors to the gateway, mean hop count to the gateway and mean number of neighbors of a mesh node were assessed in [14]. They used an SVM with a Gaussian kernel to find out the relative impact of the above topology characteristics on the aggregate throughput, fairness and node delay of the network.

2) *Fairness improvement*

Transmission Control Protocol (TCP) was first designed for wired networks and performs well over wired infrastructure; as such, when wireless networks were introduced, TCP was adopted to wireless environments. However, the fundamental differences between wireless and wired mediums result in substandard performance of TCP over wireless networks—especially affecting TCP unfairness—as TCP favors flows with smaller number of hops in WMNs. To tackle this problem, authors of [15] suggested an approach where each TCP source models the state of the system as an MDP and uses Q-learning to learn the transition probabilities of the proposed MDP based on the observed variables. To maximize TCP fairness, each node hosting a TCP source takes actions according to the recommendations of the Q-learning algorithm and adjusts TCP parameters autonomously.

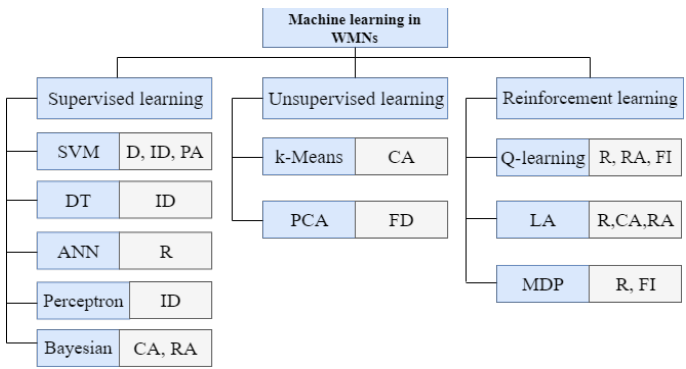

**Figure 4**. Use of different ML techniques in selected WMN applications: R=Routing, CA=Channel assignment, D=Network deployment, RA=Rate adaptation, ID=Intrusion detection, FD=Fault detection, PA=Performance analysis, FI=Fairness improvement.

IV. CONCLUSIONS AND FUTURE DIRECTIONS

We presented an overview of different design and management problems in WMNs and how different ML techniques have been applied to address them, as summarized in Fig. 4. We also discussed some issues facing current applications, presented potential solutions and highlighted ways in which ML could potentially be applied in future WMN research.

From this overview, it may be clear that RL related techniques have been responsible for a large portion of the research efforts aimed at utilizing ML in optimizing design problems of WMNs. However, even within the RL space, promising methods such as Temporal Difference (TD) learning, Dyna-Q and Prioritized sweeping have been left untouched [2]. These may provide better convergence guarantees and higher convergence rates than the already utilized methods like Q-learning and LA, which are very attractive features for a real-time stochastic system like a WMN.

Emerging techniques like Deep Learning (DL) have a huge potential towards WMN optimization. For example, in a problem like channel allocation, different metrics derived from measurable physical parameters have been used so far to evaluate effects of overlapping channels. These metrics become even less useful in scenarios where multiple neighboring networks also interfere. It is more likely that the best metrics to evaluate such complex effects is non-linear in nature and hard to be intuitively deduced. Deep Neural Networks (DNNs), which are a common implementation of DL, are ideal for these types of non-linear feature extraction tasks. As such, DNNs may be used, for example, to identify the features (metrics) that are best suited for consideration during channel assignment.

We should also make a note on Cognitive Radios (CRs). CR is a concept where the radios in a wireless network (not necessarily a WMN) intelligently manages and utilizes the limited bandwidth spectrum. Although not all intelligent WMNs have CRs on them, CR fortified WMNs, or Cognitive Radio Wireless Mesh Networks (CR-WMNs), can enable even existing intelligent optimization schemes to become better, and open the door for new ones. Therefore, future research should aim to realize the inherent symbiosis between these two technologies.